\documentclass[reprint,nofootinbib,amsmath,amssymb,fontenc aps]{revtex4-1}
\usepackage{graphicx,dcolumn,bm,hyperref,float}
\usepackage{anyfontsize}
\usepackage{color}

\newcommand{\be}{\begin{equation}}
\newcommand{\ee}{\end{equation}}
\newcommand{\ba}{\begin{eqnarray}}
\newcommand{\ea}{\end{eqnarray}}

\newcommand{\fudge}{\varepsilon}
\newcommand{\fudgephi}{\varepsilon_\phi}
\newcommand{\fudgepi}{\varepsilon_\pi}
\newcommand{\phif}{\phi_f}
\newcommand{\phit}{\phi_t}
\newcommand{\mf}{m_f}
\newcommand{\mt}{m_t}
\newcommand{\nmax}{n_{\tiny{\mbox{cut}}}}

\begin{document}

\title{Quantitative Analysis of the Stochastic Approach to Quantum Tunneling}

\author{Mark P.~Hertzberg}
\email{mark.hertzberg@tufts.edu}
\author{Fabrizio Rompineve}
\email{fabrizio.rompineve@tufts.edu}
\author{Neil Shah}
\email{neil.shah@tufts.edu}
\affiliation{Institute of Cosmology, Department of Physics and Astronomy, Tufts University, Medford, MA 02155, USA
\looseness=-1}

\begin{abstract}
Recently there has been increasing interest in alternate methods to compute quantum tunneling in field theory. Of particular interest is a stochastic approach which involves (i) sampling from the free theory Gaussian approximation to the Wigner distribution in order to obtain stochastic initial conditions for the field and momentum conjugate, then (ii) evolving under the classical field equations of motion, which leads to random bubble formation. Previous work showed parametric agreement between the logarithm of the tunneling rate in this stochastic approach and the usual instanton approximation. 
However, recent work \cite{Braden:2018tky} claimed excellent agreement between these methods. 
Here we show that this approach does not in fact match precisely; the stochastic method tends to overpredict the instanton tunneling rate. To quantify this, we parameterize the standard deviations in the initial stochastic fluctuations by $\varepsilon\,\sigma$, where $\sigma$ is the actual standard deviation of the Gaussian distribution and $\varepsilon$ is a fudge factor; $\varepsilon=1$ is the physical value. We numerically implement the stochastic approach to obtain the bubble formation rate for a range of potentials in 1+1-dimensions, finding that $\varepsilon$ always needs to be somewhat smaller than unity to suppress the otherwise much larger stochastic rates towards the instanton rates; for example, in the potential of \cite{Braden:2018tky} one needs $\varepsilon\approx1/2$. We find that a mismatch in predictions also occurs when sampling from other Wigner distributions, and in single particle quantum mechanics even when the initial quantum system is prepared in an exact Gaussian state. If the goal is to obtain agreement between the two methods, our results show that the stochastic approach would be useful if a prescription to specify optimal fudge factors for fluctuations can be developed.
\end{abstract}

\maketitle

\section{Introduction}

Quantum tunneling plays a central role in many areas of physics, including diodes, nuclear fusion, etc. In relativistic quantum mechanics of scalar bosons, it can play a central role in determining the stability, or otherwise, of vacua. In particular, in the minimal Standard Model of particle physics, the Higgs potential is only meta-stable, as it turns over at an energy scale of $\mathcal{O}(10^{11})$\,GeV, depending on the top mass \cite{EWvevmetastbl.1,EWvevmetastbl.2,EWvevmetastbl.3,EWvevmetastbl.4}. It then goes negative, rendering our electroweak vacuum meta-stable. Its lifetime from quantum tunneling has been estimated to be extremely long, much longer than the current age of the universe, at least at zero temperature. During inflation, the situation can radically change, allowing the tunneling rate to be larger \cite{Espinosa.Riotto.1,Zurek.1,Zurek.2,Joti:2017fwe,Markkanen:2018pdo,Jain:2019wxo,Fumagalli:2019ohr}. Furthermore, in the context of string theory, there are thought to be an exponentially large number of meta-stable vacua in a multi-dimensional field space. 

The computation of quantum tunneling can in principle be done by means of solving the Schr\" odinger equation directly. In the context of single particle non-relativistic quantum mechanics, this is normally a task that can be completed with fairly minimal computing resources (which we shall return to later in the paper as a testing ground). However, in the context of relativistic quantum mechanics, one must necessarily deal with a large number of degrees of freedom, which are conveniently organized into fields. In this case, the direct solution of the Schr\" odinger equation for the quantum field theory is notoriously difficult. It is then especially important to have efficient tools to provide approximate results. The most famous and celebrated approximation is to make use of instantons \cite{Coleman:1977py,Callan:1977pt}. In this method, one performs a Wick rotation on time $t\to i\,t$ to obtain the Euclideanized equations of motion. The leading instanton is the bounce solution exhibiting maximal symmetry, namely $SO(d+1)$, where $d$ is the number of spatial dimensions. 
This instanton method can be very efficient to implement and is usually thought to be an accurate approximation to the true quantum tunneling rate when the theory is in a weakly coupled regime. 

However, there are good reasons to develop alternate calculational tools. Perhaps of most interest is in the context of cosmology; in this case one is invariably studying a system with a time dependent background. This immediately spoils the $SO(d+1)$ symmetry of the Euclidean theory, so such simple instantons do not exist in this standard form. There are many possible settings where the time dependence of the background would be important: during cosmic inflation, during pre-heating after inflation, during late time dark energy, etc. In fact in the case of today's tunneling rates it is unclear if the static treatment is relevant: if the background is very slowly varying compared to the time-scale for tunneling, and one can therefore ignore the breakdown of time translational symmetry, then the tunneling is so fast our universe would have already tunneled. On the other hand, if the tunneling is sufficiently slow that we have yet to tunnel, then the background evolution matters and the static analysis is not clearly applicable. Furthermore, there are situations in which the instanton solution does not exist, while tunneling is still thought to occur, e.g., see \cite{Brown:2011ry}. So there can be many important cosmological situations in which the static instanton based method can be sub-optimal and even inaccurate. 

Therefore a much needed tool would be to have a formalism that tracks the dynamics of the tunneling process in real time rather than imaginary time. This method could then track non-trivial background dynamics that is relevant to cosmology and in other dynamical settings. This method suggests to use the real time dynamical classical equations of motion, as opposed to the imaginary time classical equations of motion which are at the heart of the usual instanton method. However, if one uses the classical equations of motion, one wonders: where does quantum mechanics enter? In such an approach, the only place it can enter is in the {\em choice of initial conditions}. 

To elaborate on this, suppose we consider a scalar field $\phi$ subject to a potential $V(\phi)$, and suppose we expand around the false vacuum $\phif$ as $V={1\over2}\mf^2(\phi-\phif)^2+\ldots$. Classically, one must specify initial conditions for the fluctuations $\delta\phi_i$ and momentum conjugate $\pi_i=\delta\dot\phi_i$. Since they are to evolve simultaneously in the classical theory, then they need to be drawn from a joint probability distribution $p(\delta\phi_i,\pi_i)$. In the true quantum theory, such a joint distribution does not exist of course; one only has a wave function $\Psi$. So instead one must choose a related distribution. A concrete proposal is to (i) approximate the wave function as the free theory ground state wave function based around the false vacuum with mass $\mf$. Since the free theory ground state wave function is Gaussian it possesses a Wigner distribution $W$ that is positive definite and can act as a joint probability distribution from which one can sample from $p=W$. By sampling from this, one can then (ii) evolve under the classical field equation equations of motion. In doing so, one finds that bubbles form randomly. This provides a novel semi-classical picture of vacuum decay, as implemented in the important work of Ref.~\cite{Braden:2018tky}, and provided the inspiration for the present study. 

By performing an ensemble average, one can show that this often leads to {\em parametric agreement with the logarithm} of the tunneling rates in the instanton approximation, and so it is can be potentially quite relevant and useful. We will refer to this as the ``stochastic approach to quantum tunneling". Pioneering work on this includes Refs.~\cite{Ellis:1990bv,Linde:1991sk}, while a more recent update that makes use of the Wigner formalism was done by some of us in Ref.~\cite{Hertzberg:2019wgx}. Other relevant work includes Refs.~\cite{Bitar:1978vx,Noorbala:2018zlv,Billam:2018pvp,Braden:2019vsw,Ai:2019fri,Blanco-Pillado:2019xny,Darme:2019ubo,Mou:2019gyl,Wang:2019hjx,Michel:2019nwa,Huang:2020bzb,Hashiba:2020rsi,Hashiba:2020rsi,Gross:2020tph}. 

It is difficult to make the case for {\em precise quantitative} agreement between this stochastic approach and the instanton
approximation in Minkowski space (non-trivial backgrounds, including inflation, are not the focus of our present study). However, in the recent work published in PRL, Ref.~\cite{Braden:2018tky} provided very interesting numerical results that appeared to drastically improve the situation. The authors implemented the stochastic method in 1+1-dimensions, by carrying out detailed numerical simulations in which the field was drawn from a Gaussian distribution, and then evolved classically. They computed the average time for bubble nucleation and deduced a tunneling rate. This was carried out in the context of a particular potential $V(\phi)$, that we will discuss later; see ahead to Eq.~(\ref{BP}). They compared this tunneling rate to that obtained from the instanton method, claiming excellent agreement. This agreement was claimed to persist as the parameters in the potential were altered and the criteria for tunneling was adjusted. 

In this paper we investigate this method carefully. We perform numerical simulations in 1+1-dimensions for a range of potentials. We also test the proposal in single particle quantum mechanics. We find that the method requires one to assume initial conditions with standard deviations {\em smaller} than those which arises from the Gaussian wave function. For the specific potential examined in Ref.~\cite{Braden:2018tky}, we find that one needs to suppress the standard deviations by a factor of $\fudge\approx1/2$. This leads to a factor of 4 change in the tail of the Gaussian distribution's exponent, which leads to a very large reduction in the tunneling rate down towards the instanton prediction; while the physical value of $\fudge=1$ leads to tunneling that is far more rapid. We show that this problem is not unique to the specific potential chosen, but afflicts general potentials. Moreover, we find that different potentials require their own different values of suppression factors $\fudge<1$ to mimic the instanton results. We also test re-scaling the fluctuations in $\delta\phi_i$ and $\pi_i$ differently, while still obeying the uncertainty principle, and find that these problems persist. Finally, we test the stochastic approach and its assumption of Gaussian initial conditions by directly comparing to a quantum problem that we numerically solve with this initial condition. Although we cannot efficiently solve this quantum problem in the context of quantum field theory, we can do so in the case of single particle non-relativistic quantum mechanics. By doing so, we find that disagreement persists.

Our paper is organized as follows:
In Section \ref{SM} we provide more details of the stochastic method.
In Section \ref{NM} we present our numerical results: a periodic potential in \ref{SecB}, a double well potential in \ref{SecDW},
addressing renormalization issues in \ref{SecRenorm},
and other physical states in Section \ref{SecOther}.
In Section \ref{QM} we study the method in single particle quantum mechanics starting with an exact Gaussian wave function.
Finally, in Section \ref{CN} we conclude.

\section{Stochastic Method}\label{SM}

In this work we will be primarily interested in standard two-derivative actions. We will focus on a single scalar field $\phi$, but will later also discuss the case of a single particle in non-relativistic quantum mechanics. The standard two-derivative action in Minkowski space is given by
\be
\mathcal{L}={1\over2}(\partial\phi)^2-V(\phi).
\ee
Here we are using the signature with $+$ for time and $-$ for space, and we work in natural units $\hbar=c=1$. Note that this can also describe some condensed matter systems if we replace $c$ by the appropriate sound speed $c_s$. 

\begin{figure}[t]
\centering
\includegraphics[width=0.97\columnwidth]{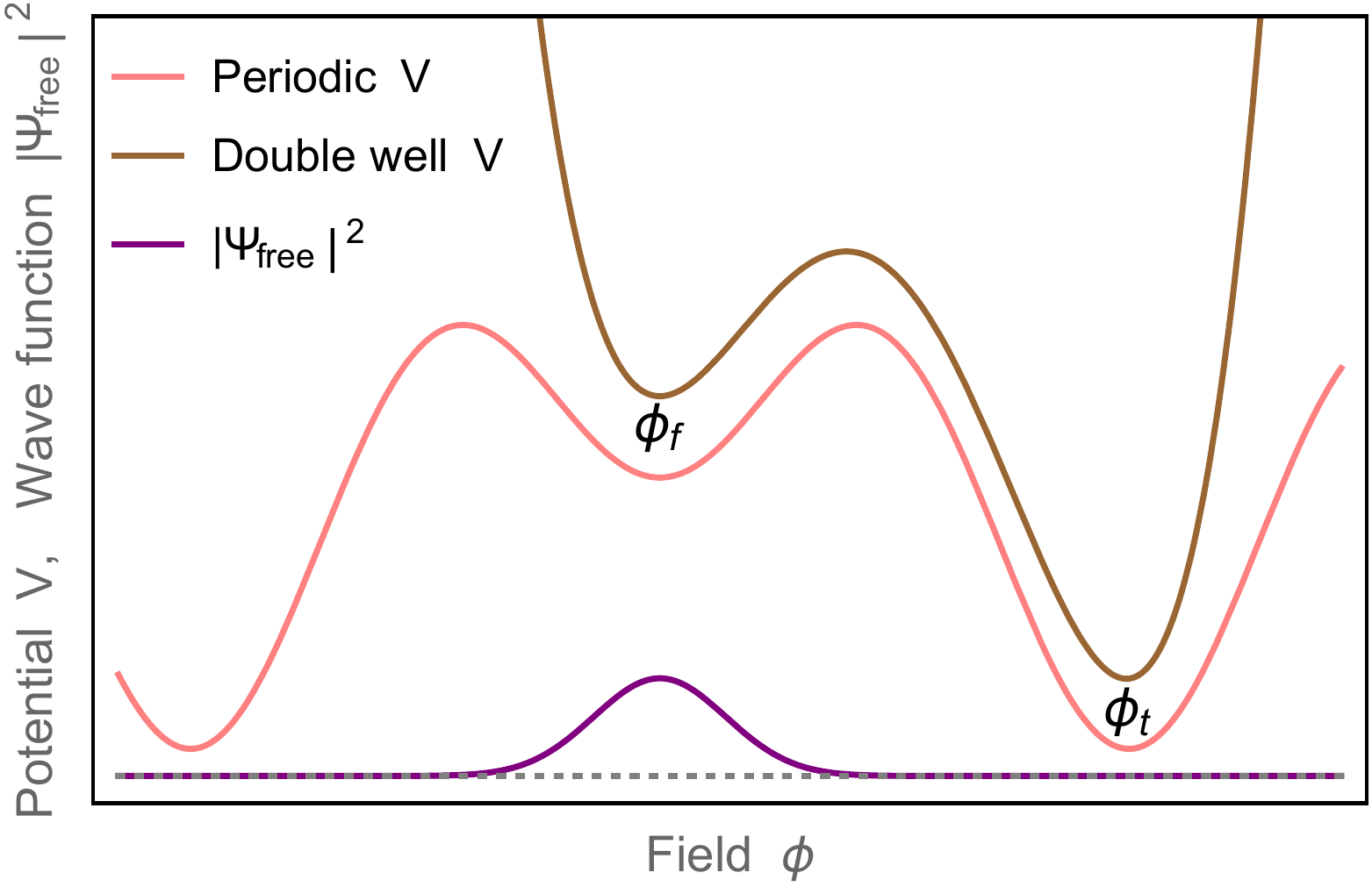}
\caption{{\em Potentials and initial condition}:
Middle pink curve: An illustrative periodic potential $V(\phi)$ of the form Eq.~(\ref{BP}). 
Upper brown curve: An illustrative double well potential $V(\phi)$ of the form Eq.~(\ref{DWP}). 
We have indicated a false vacuum by $\phif$ and a true vacuum by $\phit$. 
Lower purple curve: a cartoon of the Gaussian approximation to the wave function squared $|\Psi_{\tiny\mbox{free}}|^2$ (suppressing all the spatial dependence for ease of presentation), which is used to construct the Wigner distribution and then random initial conditions are drawn near the false vacuum in the stochastic approach.}
\label{Potential} 
\end{figure}

Consider a potential that has (at least) two local minima; one is a false vacuum at value $\phif$ and the other is a true vacuum at $\phit$, i.e., $V'(\phif)=V'(\phit)=0$ and $V(\phif)>V(\phit)$. Further, we will assume that they possess non-zero second derivatives at their respective minima, with 
\be
\mf^2\equiv V''(\phif)>0,\,\,\,\,\,\, \mt^2\equiv V''(\phit)>0.
\ee
Some illustrative example are given in Figure \ref{Potential}. Although we will focus on starting around a specific false vacuum $\phif$, in principle there can be multiple true vacua that the system can tunnel to. In a weakly coupled quantum field theory, the tunneling rate is usually thought to be well approximated by Wick rotating $t\to i t$, numerically obtaining the bounce solution, and computing the corresponding bounce action $S_B$ instanton. The tunneling rate to leading order is $\Gamma\propto e^{-S_B}$ with a pre-factor that can be determined too. 
We will return to these details later.

On the other hand, the basis of the stochastic approach is to (i) obtain stochastic initial conditions for the field and its momentum conjugate, and then (ii) evolve classically. Let us define the perturbation from the false vacuum as
\be
\phi = \phif+\delta\phi,\,\,\,\,\,\,\pi=\delta\dot\phi.
\ee
In order to obtain initial conditions, the stochastic approach normally makes use of the free theory wave functional with respect to the false vacuum. The free theory Hamiltonian is given in $d$ spatial dimensions by
\be
H_{\tiny\mbox{free}} = \int d^d x\left[{1\over2}\pi^2+{1\over2}(\nabla\delta\phi)^2+{1\over2}\mf^2(\delta\phi)^2\right].
\ee
This diagonalizes in $k$-space. The ground state wave functional is the Gaussian
\be
\Psi_{\tiny\mbox{free}}(\delta\phi)\propto \exp\!\left[-{1\over2}\int\!{d^dk\over(2\pi)^d}\omega_k|\delta\phi_{\bf k}|^2\right],\label{WF}
\ee
where $\omega_k^2=k^2+\mf^2$. Since this is a Gaussian wave function, it has a positive definite Wigner distribution
\be
W_{\tiny\mbox{free}}(\delta\phi,\pi)\propto \exp\!\left[-\!\int\!{d^dk\over(2\pi)^d}\!\left(\omega_k|\delta\phi_{\bf k}|^2+{1\over\omega_k}|\pi_{\bf k}|^2\right)\right],\label{Wigner}
\ee
The 2-point correlation functions for the field and momentum conjugate are
\ba
\langle\delta\phi_{\bf k}^* \, \delta\phi_{{\bf k}'}\rangle_{\tiny\mbox{free}} &=& {1\over 2\omega_k}(2\pi)^d\delta^d({\bf k}-{\bf k}'),\label{phik}\\
\langle\pi_{\bf k}^* \, \pi_{{\bf k}'}\rangle_{\tiny\mbox{free}} &=& {\omega_k\over 2}(2\pi)^d\delta^d({\bf k}-{\bf k}').\label{pik}
\ea
The stochastic method uses the Wigner distribution as a joint probability distribution to sample $\delta\phi$ and $\pi$ from {\em simultaneously}. This sets the variance in the field and momentum conjugate's initial conditions. By sampling from the above distribution for each $k$-mode, we obtain a representative $\phi$ back in position space. 

To define this precisely, let us consider a 1-dimensional box of spatial size $L$ in which we impose periodic boundary conditions $\phi(L/2,t)=\phi(-L/2,t)$. The $k$-modes are then discrete, obeying $k_n=2\pi n/L$, where $n$ is an integer. The system requires a cut-off on $n$ that we call $\nmax$. By sampling from the wave functional, we obtain the following stochastic initial conditions
\ba
\delta\phi_i(x) &=& {1\over\sqrt{L}}\sum_{n=1}^{\nmax} e^{ik_nx}\phi_n+c.c,\label{phii}\\
\pi_i(x) &=& {1\over\sqrt{L}}\sum_{n=1}^{\nmax} e^{ik_nx}\pi_n+c.c,\label{pii}
\ea
where the factor $1/\sqrt{L}$ is a useful pre-factor when switching from the continuum to the discrete in a box of size $L$. Here the coefficients $(\phi_n,\,\pi_n)$ are random complex numbers; they have uniformly distributed phase on the domain $[0,2\pi)$ and magnitude drawn from a Gaussian distribution, with standard deviation given by
\ba
\Delta\phi_n\equiv\sqrt{\langle|\phi_n|^2\rangle} &=&\fudgephi{1\over\sqrt{2 \omega_n}},\label{phin}\\
\Delta\pi_n\equiv\sqrt{\langle|\pi_n|^2\rangle} &=&\fudgepi\sqrt{\omega_n\over2}.\label{pin}
\ea
Here the factors of $1/\sqrt{2\omega_n},\,\sqrt{\omega_n/2}$ comes from enforcing the $k$-space standard deviations of Eqs.~(\ref{phik},\ref{pik}). 

Importantly, we have also included a pair of ``fudge factors" $\fudgephi,\,\fudgepi$ in our expressions for the standard deviations in Eqs.~(\ref{phin},\ref{pin}). The correct value to obtain the actual fluctuations of the free theory of Eqs.~(\ref{WF}-\ref{pik}) is 
\be
\fudgephi=\fudgepi = 1\,\,\,\,\,\,\mbox{(physical value)}.
\ee
If we re-scale the fluctuations in $\delta\phi_i$ and $\pi_i$, then the value assigned to the standard deviations can violate the uncertainty principle. One now has
\be
\Delta \phi_k\,\Delta\pi_k={\fudgephi\fudgepi\over2},\label{Heisenberg}
\ee
rather than the usual value of just $1/2$ (recall that we are setting $\hbar=1$). If we set $\fudgepi=1/\fudgephi$ then we are still saturating the uncertainty principle; these correspond to physical states in the Hilbert space and and we will return to them in Section \ref{SecOther}. 

As we will see, it will be convenient to include the possibility of standard deviations that are suppressed relative to the physical value by these factors $\fudgephi,\,\fudgepi$. As an example, one of the potentials we will be interested is the upcoming Eq.~(\ref{BP}), which was the focus of the paper \cite{Braden:2018tky}. In this case we will find that we need to use the fudge factors of 
\be
\fudgephi=\fudgepi \approx {1\over2}\,\,\,\,\,\,\mbox{(Optimal value for Eq.~(\ref{BP})}),
\ee
in order to obtain reasonable quantitative agreement with the instanton approximation. Such values are violating the uncertainty principle by a factor of 1/4 and cannot be associated with a Wigner distribution. Now it is important to note that a change in the $\fudge$'s by any $\mathcal{O}(1)$ number, changes the distribution dramatically, as we will see quantitatively in the next Section. If one re-scales the standard deviation by a factor of $\fudge=1/2$, then one is enhancing the coefficient in the exponent of the Gaussian by a factor of 4 in $\delta\phi_i$ and $\pi_i$. Since one is often interested in situations at small coupling, where tunneling is rare, then one is deep in the {\em tail} of the Gaussian in order to find a configuration that can organize into a bubble and tunnel. This means that a factor of a few changes in the exponent (in either $\delta\phi_i$ or $\pi_i$) can translate into a huge change in the tunneling rate, as we shall see.

In this work we will see that the physical value of $\fudgephi=\fudgepi=1$ will produce tunneling rates that are much larger than the corresponding rates from the instanton method, and one will indeed need lower values of the $\fudge$'s. We will explore different potentials and different dimensionality (both $d=1$ and $d=0$) to see how $\fudge$ needs to change to force the stochastic method to show better agreement.

In any case, once the initial conditions are specified, then they are evolved under the classical field theory equations of motion
\be
\ddot\phi-\nabla^2\phi+V'(\phi)=0.
\ee
Due to the non-trivial initial conditions, there can occasionally be a sufficient amount of energy accumulated in a region of space for the field to climb over the barrier, allowing a bubble to form and then expand; this effectively mimics the idea of ``tunneling". This simulation can be repeated numerous times for an ensemble average to determine a tunneling rate. Then one can compare to the familiar instanton rate, which in this context is usually thought to accurately describe the quantum tunneling rate in the weak coupling regime. In the next Section we will carry out this numerical procedure in detail.

\section{Potentials and Numerical Results}\label{NM}

In this section we will numerically explore several potentials $V(\phi)$. We will then explore renormalization issues and multiple choices for the initial fluctuations. 

\subsection{Periodic Potential}\label{SecB}

As a starting point, we consider a periodic potential that was the primary focus of Ref.~\cite{Braden:2018tky}, namely
\be \label{BP}
V(\phi) = V_0\left(-\cos\!\left(\phi\over\phi_0\right)+{\lambda^2\over2}\sin^2\!\left(\phi\over\phi_0\right)\right).
\ee
Here $V_0,\,\phi_0,\,\lambda$ are parameters. Formally, one can scale both $V_0$ and $\phi_0$ out of the classical equations of motion, leaving just 1 dimensionless parameter $\lambda$, which controls the relative heights of vacua. However, in such re-scaled variables, the value of $\phi_0$ will reappear in the initial conditions, just as it will appear in the quantum problem. 

This potential is periodic and so it has an infinite number of false and true vacua. For $\lambda>1$ it has meta-stable false vacua. An illustrative plot is given as the middle pink curve in Figure \ref{Potential}. We can focus on the false vacuum at $\phif=\pi\,\phi_0$ and determine the false mass $\mf$ by expanding around this point. A convenient aspect of this potential is that one can use $\cos(\phi/\phi_0)$ as a kind of diagnostic as to whether one is close to the false vacuum or the true vacuum. The value of $\phi=\phi(x)$ varies from place to place, so it is useful to perform its spatial average $\langle\cos(\phi/\phi_0)\rangle$. A value near $-1$ means that the field is close to the false vacuum, while a value close to $+1$ means that the field is close to the true vacuum.

\begin{figure}[t]
\centering
\includegraphics[width=\columnwidth]{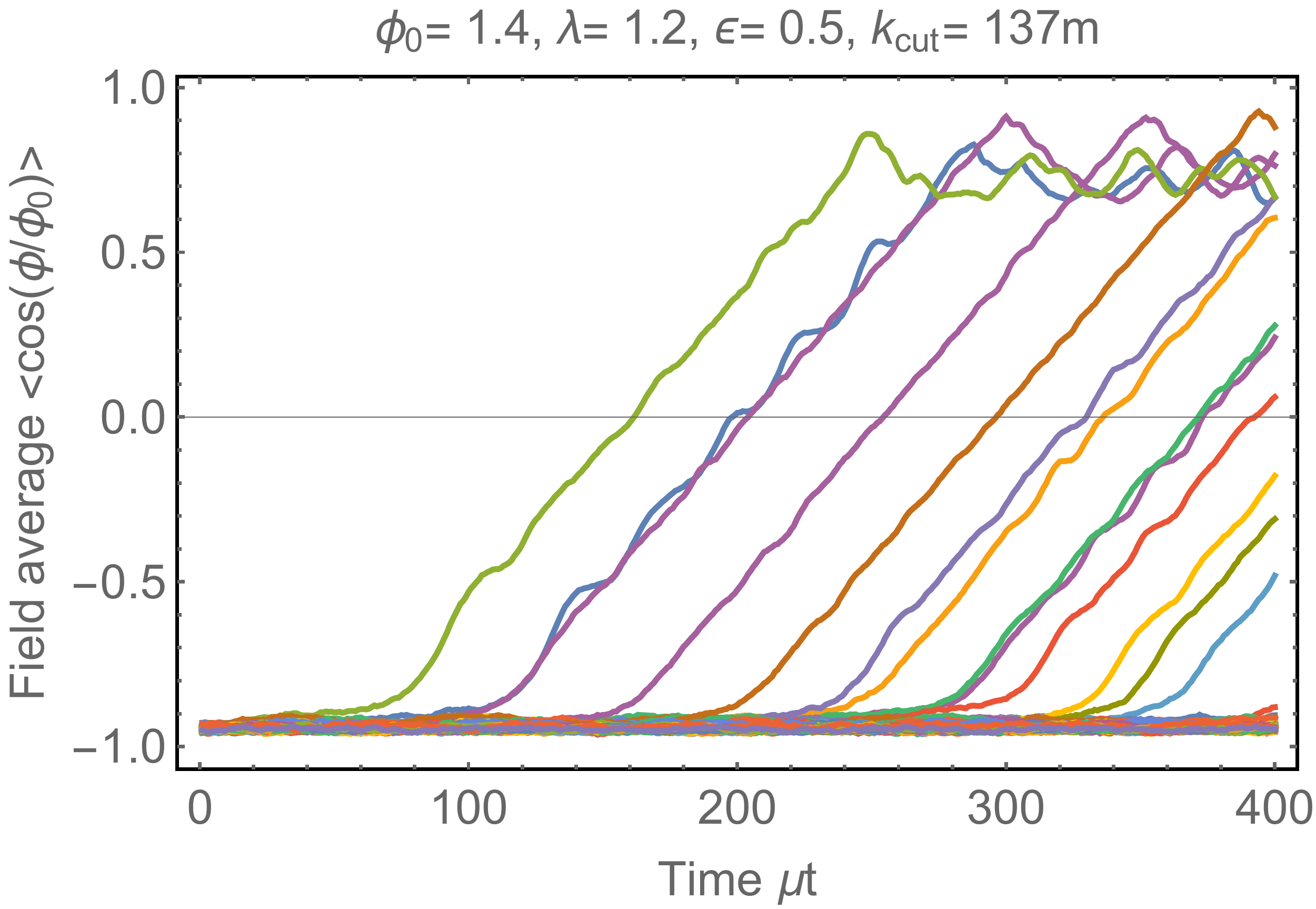}\\
\vspace{0.3cm}
\includegraphics[width=\columnwidth]{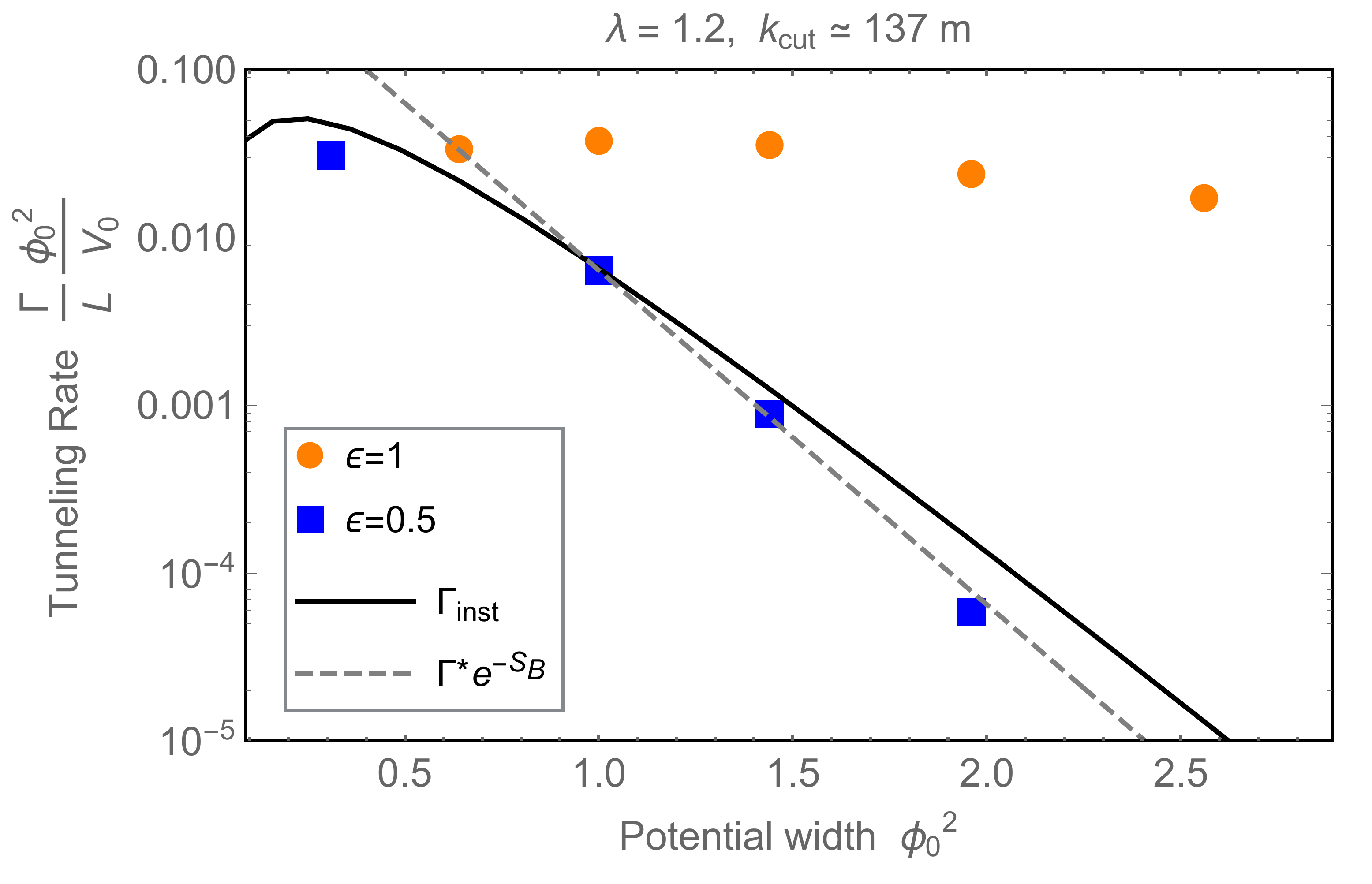}
\caption{{\em The periodic potential Eq.~(\ref{BP})}. 
Top panel: The spatial average $\langle\cos(\phi/\phi_0)\rangle$ versus time for a set of random realizations in this model. The chosen potential width parameter is $\phi_0=1.4$ and we used the fudge factor of $\fudge=1/2$ to suppress the fluctuations.
Bottom panel: The normalized tunneling rate $\Gamma$ versus the effective width of the potential $\phi_0$ for two different choices of $\fudge$, namely $\fudge=1$ (orange dots) and $\fudge=0.5$ (blue dots). 
Tunneling rates have been computed with ensembles of 200 realizations.
In these figures we have chosen $L=400/\mu, V_0/\phi_0^2=0.008\mu^2$, $\lambda=1.2$, and $k_{\text{cut}}=k_{\star}$, similar to Ref.~\cite{Braden:2018tky} (in practice we fix $n_{\text{cut}}=518$).}
\label{CosAveB} 
\label{RateB} 
\end{figure}

We drew our initial conditions from the above Gaussian distributions, in accord with the stochastic method, and solved the nonlinear classical field equation of motion. This was done implementing a standard leap-frog integrator in Python as well as in Mathematica. We checked that in our system the total energy changed by a small percentage on the order of $(0.1-0.01)\%$, and the errors in the field evolution were small as well. We have considered two different choices of cutoff in Eq.~(\ref{phii}) and Eq.~(\ref{pii}). First, we fixed $k_{\text{cut}}$ to the value $k_{\text{cut}}=k_{\star}=(2\pi/L) n_{\star}=1024\pi/(25\sqrt{2})\sqrt{V_0}/\phi_0$ similar to Ref.~\cite{Braden:2018tky}. However, this cutoff is in general so large that  renormalization effects can become relevant. For this reason, we also considered lower cutoffs such as to be in a regime where renormalization is negligible, which we will discuss shortly. The results for the spatial average $\langle\cos(\phi/\phi_0)\rangle$ for many different random realizations (200 in this example) as a function of time are given in Figure \ref{CosAveB} (top panel), with cutoff given by $k_\star$. The figure shows very distinctive behavior; for a while (on the order of $(100-200)/\mu$ in this example) the system is just fluctuating around the false vacuum. Then at some moment a sufficiently large fluctuation allows a bubble to form. The bubble expands out at approximately the speed of light. It eventually fills the entire box and one has completed the tunneling transition to the true vacuum. We have chosen parameters for the potential $\phi_0=1.4$ and $\lambda=1.2$ for this plot. However, qualitatively similar behavior occurs for other parameters. 

We extract a tunneling rate from the ensemble by examining the timescale over which the classical fields ``tunnel'', using the prescription outlined in \cite{Braden:2018tky}. We declare a field as ``tunneled'' if its value of $\langle\cos(\phi/\phi_0)\rangle$ averaged over the box exceeds a threshold which is determined as follows. We define two terms $\bar{c}_T$ and $\Delta c_T$, which are the ensemble average and standard deviations respectively of $\langle\cos(\phi/\phi_0)\rangle$ over the box at $t=0$. The threshold is then defined as $\bar{c}_T + n_\sigma \Delta c_T$, where $n_\sigma$ is some constant. Our results are consistent with the claim in \cite{Braden:2018tky} that the tunneling rate is invariant to the choice of $n_\sigma$ between 5 and 25, so we choose its value to be 15 for our simulations. We show the tunneling rates versus the potential width $\phi_0^2$ in Figure \ref{RateB} (bottom panel) for an example set of parameters $V_0/\phi_0^2=0.008\mu^2$ and $\lambda=1.2$, consistent with those chosen for an analogous figure in \cite{Braden:2018tky} ($\mu$ just sets units; one can set units $\mu=1$ everywhere in this paper).

Since we cannot solve the full quantum field theory efficiently, we compare these results to the instanton approximation. In Figure \ref{RateB} (bottom panel) the solid black line is the instanton result, while the dashed grey line is from just the bounce action, i.e. $\Gamma^{*}e^{-S_B}$ with $\Gamma^{*}$ a constant numerical prefactor which is fixed by imposing agreement with the numerical result for $\fudge=0.5, \phi_0=1$. Recall that to obtain the bounce action, one begins by obtaining solutions to the Euclidean equations of motion with $SO(d+1)$ symmetry
\be
\phi''+{d\over r}\phi'=V'(\phi),
\ee
with boundary conditions $\phi'\to0$ as $r\to0$ and $\phi\to\phif$ as $r\to\infty$. The corresponding bounce action is
\be
S_B = A_{d} \int dr\,r^{d}\left[{1\over2}(\phi')^2+V(\phi)-V(\phif)\right],
\ee
where $A_{d}$ is the area of the $d$-dimensional unit sphere. The leading approximation for the tunneling rate is 
\be
\Gamma_{\text{inst}}=\Gamma_0 \left(\frac{S_B}{2\pi}\right)^{d+1}e^{-S_B},
\ee
where $\Gamma_0$ is a determinant prefactor which depends on the parameters of the potential. It is straightforward to compute these bounce actions in these simple field theories. In this paper, we neglect renormalization effects in the determinant prefactor. However, we will discuss the issue of loop corrections to the mass in Section \ref{SecRenorm}. In the figures we use the approximation $\Gamma_0=N m^2 L$, where $N$ is the number of nearby true minima (so $N=2$ in (\ref{BP}) and $N=1$ in the upcoming (\ref{DWP})). 

In the figure we have compared the instanton rates to the stochastic method with $\fudge=\fudgepi=\fudgephi$. For the latter, we have included both the $\fudge=1$ (orange points) and the $\fudge=1/2$ (blue points) choices for the initial fluctuations. We see that the physical value of $\fudge=1$ greatly overpredicts the tunneling rate, especially at large $\phi_0$, which is the weak coupling regime in which the instanton is usually thought to be the most trustworthy. On the other hand, by suppressing the fluctuations with $\fudge=1/2$ (or $\fudge\approx1/2$), there is observed to be rather good agreement between the two methods. The values in this latter curve (blue) are seen to be similar to the results in Fig.~2 of \cite{Braden:2018tky}.

\subsection{Double Well Potential}\label{SecDW}

\begin{figure}[t]
\centering
\includegraphics[width=\columnwidth]{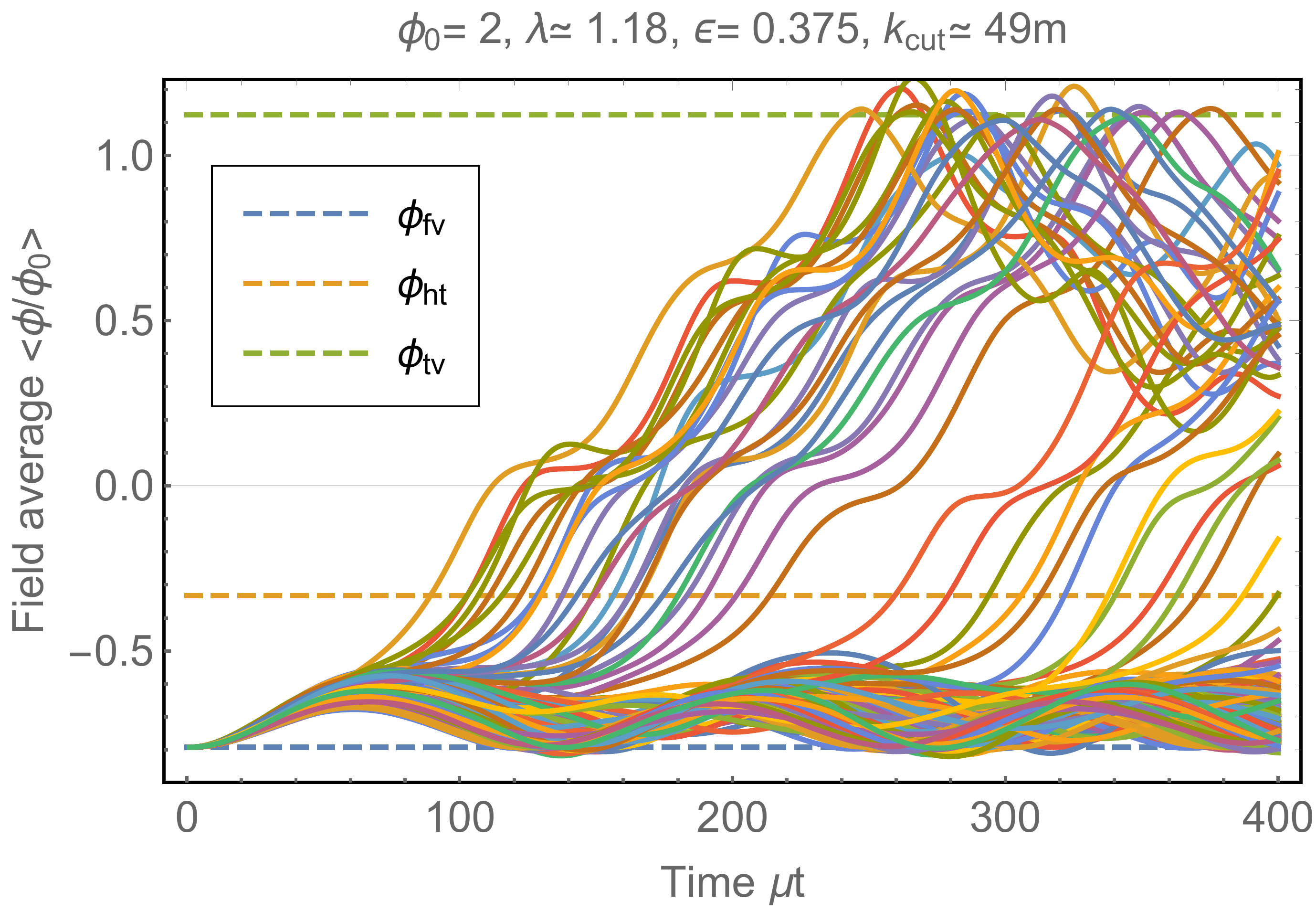}\\
\vspace{0.3cm}
\includegraphics[width=\columnwidth]{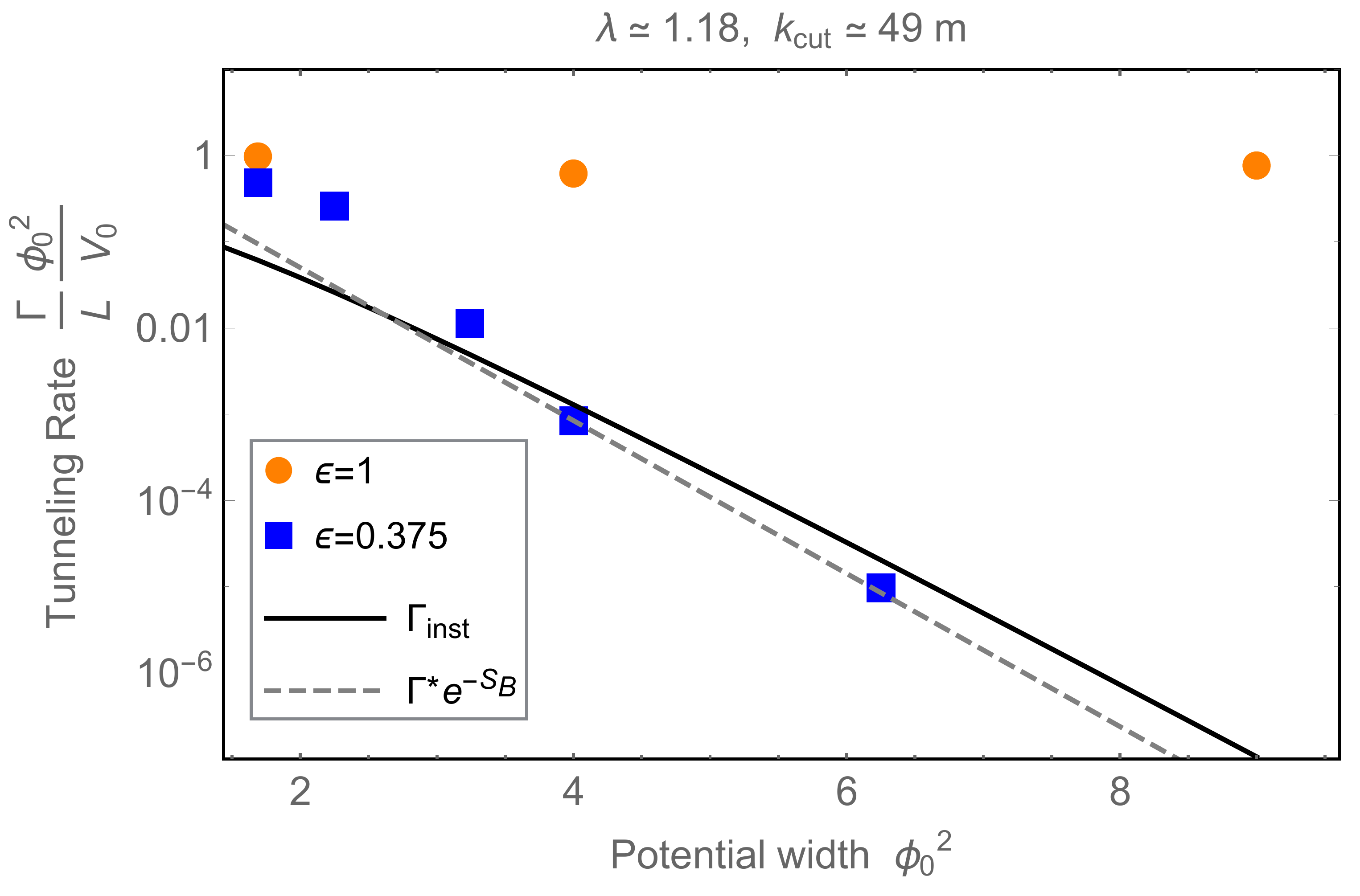}
\caption{{\em The double well potential Eq.~(\ref{DWP})}: 
Top panel: The spatial average $\langle\phi/\phi_0\rangle$ versus time for a set of random realizations in this model. Here $\phi_{\text{fv}},\,\phi_{\text{ht}},\,\phi_{\text{tv}}$ are the locations of the false vacuum, the hilltop, and the true vacuum, respectively. The chosen potential width parameter is $\phi_0=2$ and we used the fudge factor $\fudge=0.375$ to suppress the fluctuations.
Bottom panel: The normalized tunneling rate $\Gamma$ versus the effective width of the potential $\phi_0$ for two different choices of $\fudge$, namely $\fudge=1$ (orange dots) and $\fudge=0.375$ (blue dots). 
Tunneling rates have been computed with ensembles of 300-3000 realizations depending on the value of $\phi_0$.
In these figures we have chosen $L=400/\mu$, $V_0/\phi_0^2=0.001\mu^2$, $\lambda\approx1.18$, and $k_{\text{cut}}= k_{\star}$ (in practice we fix $n_{\text{cut}}=183$).}
\label{DWav} 
\label{RateDW} 
\end{figure}

We will also consider the case of a double well potential. In particular, we consider the following quartic potential
\be
V(\phi) = V_0\left(\left(1-{\phi^2\over\phi_0^2}\right)^{\!2}+\lambda\left(1-{\phi\over\phi_0}\right)\right),
\label{DWP}\ee
where $\lambda$ is another dimensionless parameter that controls the energy separation between the vacua. An illustrative plot is given as the upper brown curve in Figure \ref{Potential}. In this case, we followed the spatial average of the field, which is shown in Figure \ref{DWav} (top panel) for a representative choice of parameters. We then declare that a realization has tunneled once $\langle\phi/\phi_0\rangle$ is larger than the location of the maximum of the barrier which separates the false and true minima (the precise choice of threshold does not affect our results). 

The tunneling rates from simulations using the stochastic method and the instanton are given in Figure \ref{RateDW} (bottom panel), fixing $k_{\text{cut}}= k_{\star}$ (again with $\fudge=\fudgephi=\fudgepi$). Here we again see that the physical value of $\fudge=1$ overpredicts the instanton tunneling. In this case, however, we find that the fudge factor of $\fudge\approx1/2$ is not the optimal choice, as it was in the previous periodic potential. We find that for this model with $\lambda\approx1.18$, the fudge factor $\fudge\approx0.38$ provides the best fit. 

This implies that there is no universal value for $\fudge$, the amount by which one should suppress fluctuations in order to reproduce the instanton tunneling rates.

\subsection{Renormalization and Lower Cutoffs}\label{SecRenorm}

Let us now discuss renormalization effects. The one-loop quantum correction to the mass in 1+1-dimensions in both cases above is given by \cite{Coleman:1973jx}
\begin{equation}
m_{R}^2= m_B^2+\frac{g}{2}\frac{1}{4\pi}\log\left(\frac{k_\text{cut}^2+m_B^2}{m_B^2}\right),
\end{equation}
where $m_B^2$ is the bare mass computed as $V''(\phi=\phi_{\text{fv}})$ for the potentials Eqs. (\ref{BP}), (\ref{DWP}) defined at the cutoff $k_\text{cut}$. While
$g$ is a coefficient given by $g=V_0/\phi_0^4(1-4\lambda^2)$ for the potential Eq. (\ref{BP}) and $g=24V_0/\phi_0^4$ for the potential  Eq. (\ref{DWP}). By imposing that the renormalized mass is not very different from the bare mass $|m_R^2-m_B^2| < |m_B^2|$, so that renormalization effects are small, one can find an upper bound on $k_\text{cut}$. The choice $k_{\text{cut}}=k_\star$, which was considered above to compare with Ref.~\cite{Braden:2018tky}, turns out to be above this upper bound for the values of parameters considered here.

\begin{figure}[t]
\centering
\includegraphics[width=\columnwidth]{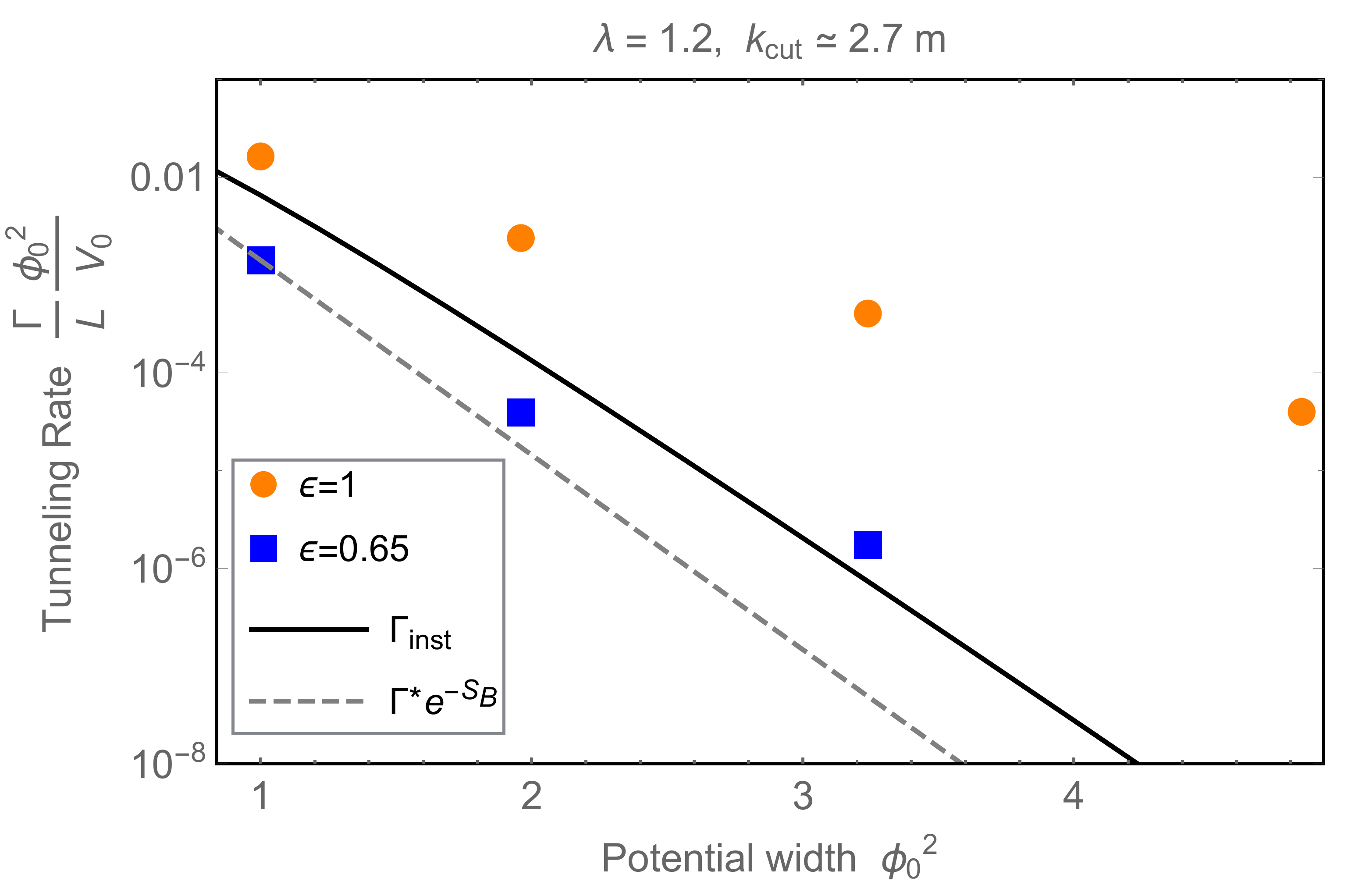}\\
\vspace{0.3cm}
\includegraphics[width=\columnwidth]{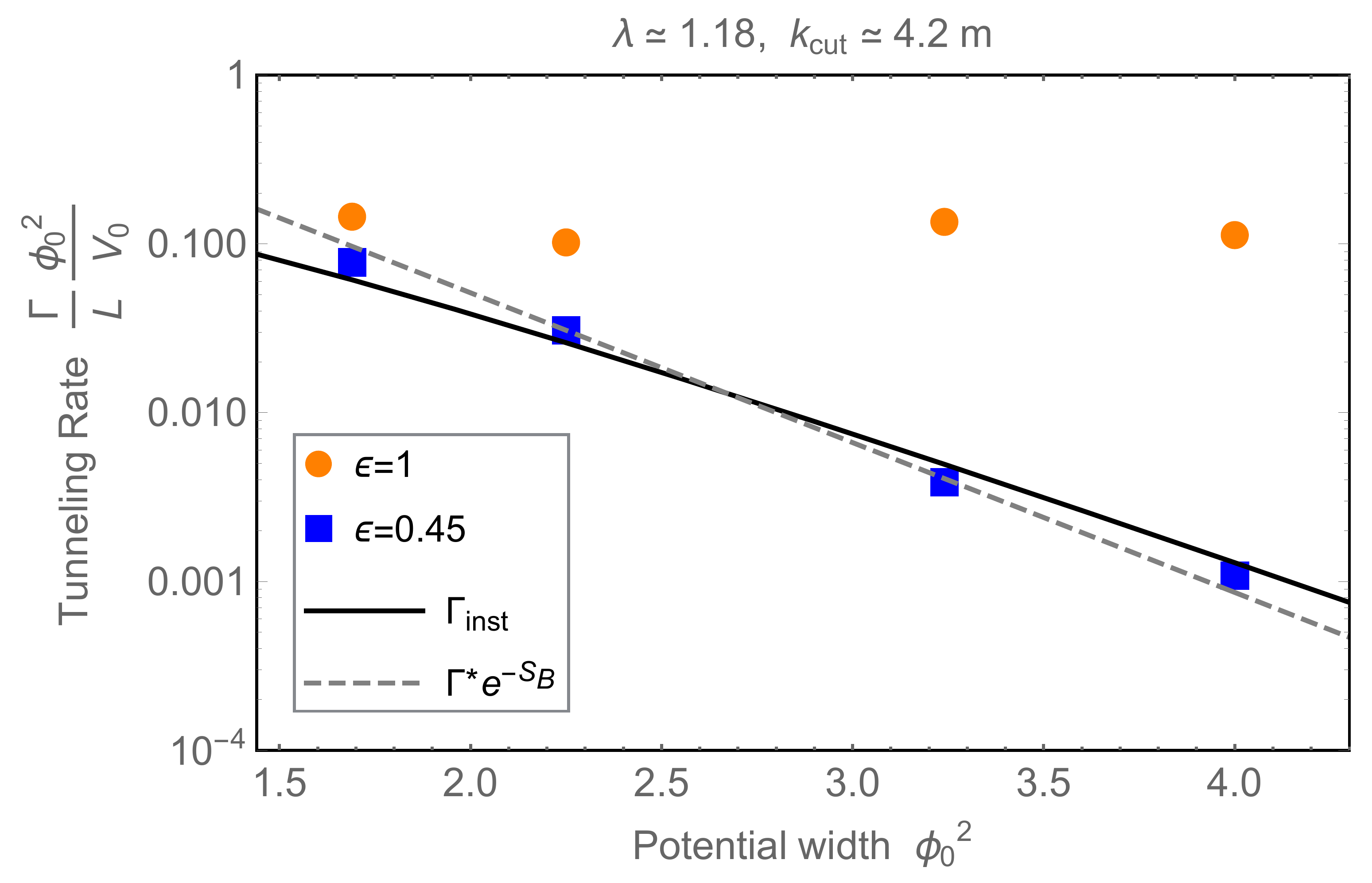}
\caption{{\em Suppressing renormalization effects}: 
Top panel: The normalized tunneling rate $\Gamma$ versus the effective width of the potential $\phi_0$, in the period potential of Eq.~(\ref{BP}), for two different choices of $\fudge$, namely $\fudge=1$ (orange dots) and $\fudge=0.65$ (blue dots). 
We have chosen $L=400/\mu$, $V_0/\phi_0^2=0.008\mu^2$, $\lambda=1.2$, and $n_{\text{cut}}=10$. Tunneling rates have been computed with an ensemble of 1000-5000 realizations depending on the value of $\phi_0$.
Bottom panel: The normalized tunneling rate $\Gamma$ versus the effective width of the potential $\phi_0$, in the double well potential of Eq.~(\ref{DWP}), for two different choices of $\fudge$, namely $\fudge=1$ (orange dots) and $\fudge=0.45$ (blue dots). 
We have chosen $L=400/\mu$, $V_0/\phi_0^2=0.001\mu^2$, $\lambda\approx1.18$, and $n_{\text{cut}}=16$. Tunneling rates have been computed with an ensemble of 500-3000 realizations depending on the value of $\phi_0$.}
\label{RateB2} 
\label{RateDW2} 
\end{figure}

One may then wonder if better agreement between the stochastic approach and the instanton approximation can be obtained for a lower choice of cutoff to reduce any renormalization effects. We have considered this possibility and fixed $n_{\text{cut}}=10$ for the potential of Eq.~(\ref{BP}) and $n_\text{cut}=16$ for the potential of Eq.~(\ref{DWP}). For these choices, we have checked that renormalization is negligible and a nucleating bubble is still well resolved. The results are presented in Figure \ref{RateB2} for both the physical choice $\fudge=1$ as well as other choices of fudge factors that give rough agreement between the two approaches (again with $\fudge=\fudgephi=\fudgepi$). Once again, we observe that the physical choice $\fudge=1$ leads to an overestimate of the instanton rate of tunneling. 

We find that the optimal fudge factors differ from the ones used with $k_{\text{cut}}= k_{\star}$. This indicates that the optimal fudge factor is cutoff-dependent in this regime.

\subsection{Other Physical Initial States}\label{SecOther}

Finally, one can consider a class of fudge factors which arises from physical states, while being different from the most physically motivated value $\fudgephi=\fudgepi=1$. By re-scaling the argument of the wave-function in Eq.~(\ref{WF}) by a factor of $1/\fudge^2$, we obtain the following Wigner distribution
\be
W(\delta\phi,\pi)\propto \exp\!\left[-\!\int\!{d^dk\over(2\pi)^d}\!\left({\omega_k\over\fudge^2}|\delta\phi_{\bf k}|^2+{\fudge^2\over\omega_k}|\pi_{\bf k}|^2\right)\right].\label{WignerFudge}
\ee
For general $\fudge\neq1$, this no longer corresponds to the Wigner distribution of the free quadratic theory, and so may not seem an especially well motivated initial condition. However, it is a physical state in the Hilbert space, and saturates the uncertainty principle with $\fudge=\fudge_\phi=1/\fudgepi$  (recall Eq.~(\ref{Heisenberg})). This is to be contrasted with the fudge factors used in the previous subsections with $\fudge=\fudgephi=\fudgepi$. 

\begin{figure}[t]
\centering
\includegraphics[width=1.03\columnwidth]{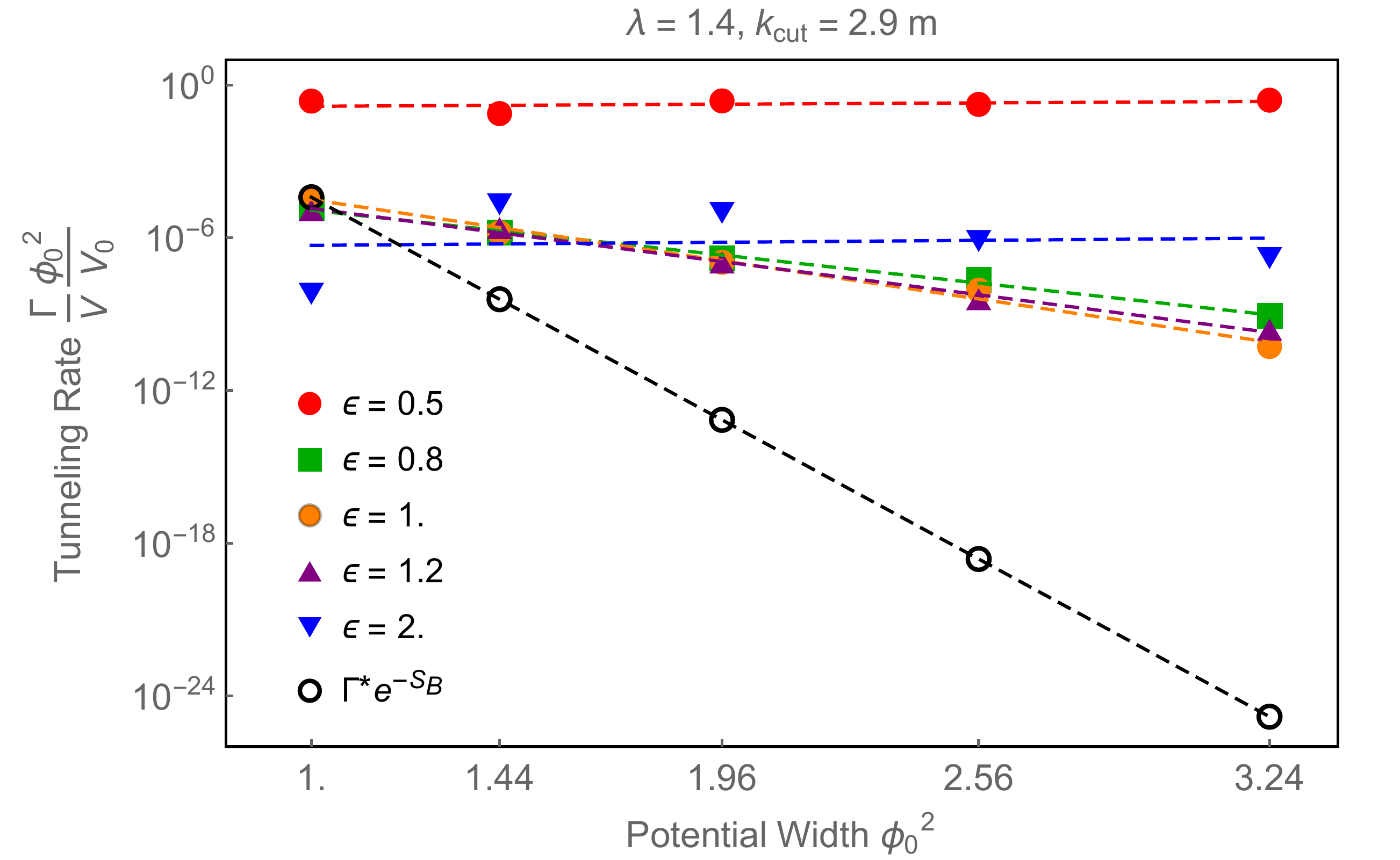}
\caption{{\em Physically allowed initial state fluctuations}:
The normalized tunneling rate $\Gamma$ versus the effective width of the potential $\phi_0$ in the periodic potential of Eq.~(\ref{BP}). Stochastic initial conditions realizations have been obtained with the following choices of fluctuations which all saturate the uncertainty principle $\fudge\equiv\fudgephi=1/\fudgepi$. We have chosen $L=400/\mu$, $V_0/\phi_0^2=0.008\mu^2$, $\lambda=1.4$, and $n_{\text{cut}}=16$ to avoid renormalization issues. Tunneling rates have been computed with an ensemble size of 400 for all $\phi _0$.}
\label{RateBPInverse} 
\end{figure}

We have considered this family of options for initial fluctuations in the stochastic method and applied them to the potential of Eq.~(\ref{BP}). The results are given in Figure \ref{RateBPInverse}. For comparison we still compare to the standard instanton result. (One could modify the quantum theory by altering its initial condition accordingly, but that is not our focus right here.) Once again, we observe that for any choice within this family of physically allowed choices for initial fluctuations, we see a dramatic overprediction of the instanton tunneling rates at large $\phi_0$. In particular we see that if $\fudge$ is small, we can no longer suppress the rates towards the instanton values (approximated by the exponent of bounce action in dashed-black curve), because to satisfy the Heisenberg uncertainty principle, the Wigner distribution implies a corresponding large value for $\pi_i$, causing the field to have so much kinetic energy it easily goes over the hill-top and tunnel quickly. Similarly if $\fudge$ is large, we suppress $\pi_i$, but enhance $\delta\phi_i$ so much it readily tunnels. In fact we see that $\fudge=1$ is roughly the best choice (see orange data points) among this space of physically allowed initial conditions.

\section{Initial Gaussian Wave Function in Quantum Mechanics}\label{QM}

Having established in the previous section that the stochastic approach does not quantitatively match the instanton results, unless one artificially reduces the amplitude of fluctuations, an important issue is to identify why and to test the stochastic approach directly. Recall that there are 2 aspects to the stochastic method: (i) drawing initial conditions for $\delta\phi_i$ and $\pi_i$ from the free theory Gaussian Wigner distribution, and (ii) evolving under the classical equations of motion. In this section, we will test the assumption (i) in a direct way. To do this, we will promote (i) to a fact of the quantum theory, i.e., we set the quantum initial condition to an actual Gaussian wave function. We then solve the Schr\" odinger equation directly and compare to the stochastic approach. In fact one might consider the initial Gaussian wave function to be the standard prescription in a regular quantum mechanical tunneling problem (as some reasonable initial condition does need to be selected after all); we would like to evolve this system precisely and compare. 

\begin{figure}[t]
\centering
\includegraphics[width=0.94\columnwidth]{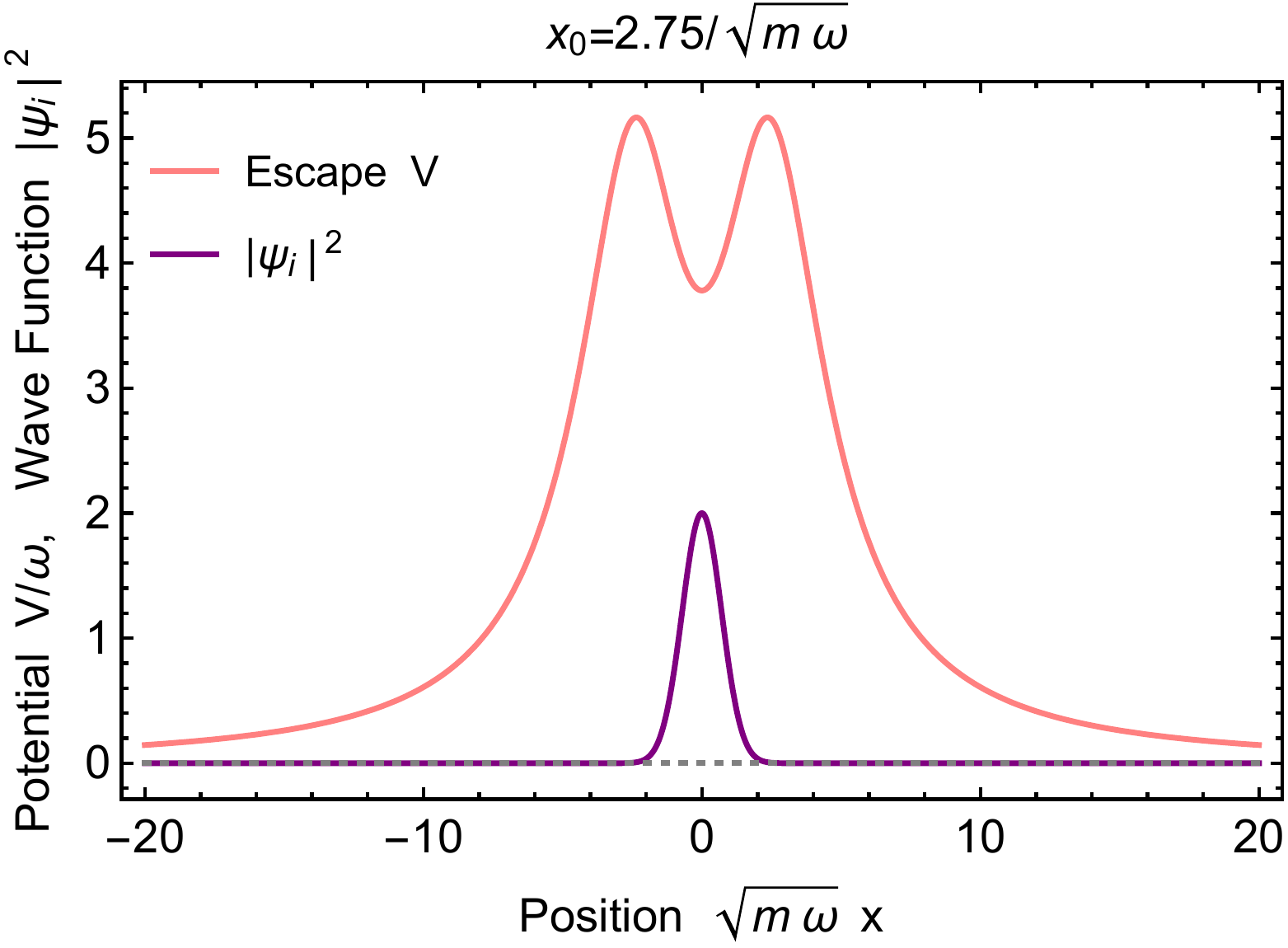}\!\!\!\!\!\!\\
\vspace{0.3cm}
\includegraphics[width=\columnwidth]{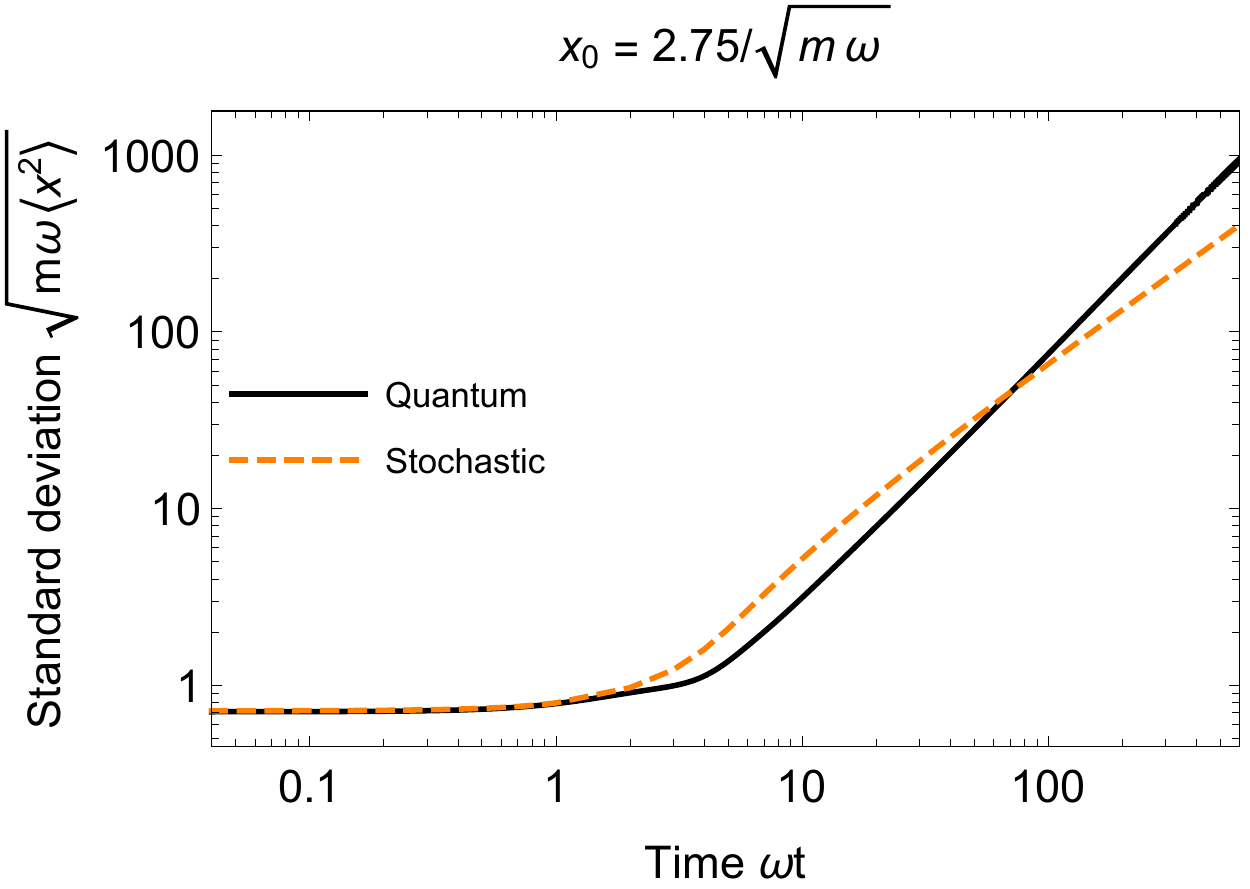}
\caption{{\em Single particle quantum mechanics}:
Top panel: The potential (pink curve) used in this section to investigate tunneling in quantum mechanics and (un-normalized) initial Gaussian wave function squared $|\Psi_i|^2$ (purple curve). The potential is given in Eq.~(\ref{PE}) with parameter $x_0=2.75/\sqrt{m\omega}$ and the initial Gaussian wave function is given in Eq.~(\ref{PsiIni}).
Bottom panel: The standard deviation in position versus time. The solid-black curve is the full numerical solution of the Schr\" odinger equation. The dashed-orange curve is the corresponding solution from the stochastic method. We performed the simulation in a box of size $L=3536/\sqrt{m\omega}$.}
\label{QMPot} 
\label{QMescape} 
\end{figure}

Since solving the Schr\" odinger equation directly in quantum field theory is so difficult (and is the primary reason to develop alternate approximation schemes in the first place), we turn to single particle non-relativistic quantum mechanics. This provides a useful, albeit limited, testing ground for the method. As a quantum mechanics analogue of the quantum field theory tunneling process, we focus on the {\em escape} of a single particle in the following potential
\be
V(x)={1\over2}{m\omega^2 x^2}\frac{1-{1\over2} \left(\frac{x}{x_0}\right)^2}{1+{1\over2} \left(\frac{x}{x_0}\right)^4},
\label{PE}\ee
which is initially located near $x=0$. This is shown in Figure \ref{QMPot} (top panel), shifted up by a constant. Note that this has only one local minimum at $x=0$.\footnote{We chose Eq. \ref{PE} instead of a double well potential because in quantum mechanics the position of a single particle initially sitting in one of the wells oscillates in time between the two wells. Therefore, in this case one cannot really talk of particle escape.} The tunneling rate for this potential in quantum mechanics is known to go as $\Gamma\propto e^{-S_{\tiny{\mbox{WKB}}}}$~\cite{Coleman:1985rnk}, where $S_{\tiny{\mbox{WKB}}}$ is the WKB factor.

For a precise comparison between the quantum and stochastic method, we now choose to start the quantum state in the exact Gaussian approximation to the wave function (defined with respect to the quadratic minimum of the potential)
\be
\psi_i(x) = \left(m\omega\over\pi\right)^{\!1/4} \exp\!\left[-{1\over2}m\omega\,x^2\right].
\label{PsiIni}\ee
(this is also shown in Figure \ref{QMPot} (top panel)).
We then solve the time dependent Schr\" odinger equation numerically in the full potential $V(x)$. 

The initial quantum state will evolve. Over time the wave function spreads out towards infinity, which corresponds to the particle escaping from the well. We track the variance in position over time 
\be
\langle x^2 \rangle_Q(t) = \int_{-\infty}^\infty dx\,|\psi(x,t)|^2 \,x^2.
\ee
For comparison, we can again perform the stochastic analysis to obtain an approximation to the variance. We do so as follows: First, we create an ensemble of $10^4$ initial conditions for position and momenta chosen from Gaussian distributions whose variances are sourced from the quantum variances, so 
\be
\sigma_{x,i}^2 = \frac{1}{2 m\omega}, \,\,\,\,\,\, \sigma_{p,i}^2 = \frac{m\omega}{2},
\ee
(i.e., $\fudge_x=\fudge_p=1$). We then evolve each member of the ensemble using the classical equations of motion and compute the ensemble average $\langle x^2 \rangle_S (t)$. 

The numerical results for the quantum and stochastic values are given in Figure \ref{QMescape} (bottom panel), with $x_0=2.75/\sqrt{m\omega}$. We have chosen a box size of $L=3536/\sqrt{m\omega}$ and checked that this is large enough to ensure boundary reflection effects are small for the times plotted in the figure. We see that, while both approaches are somewhat similar for this choice of parameter, there are important differences: the stochastic method (dashed-orange) rises above the quantum result (solid-black), then goes lower than it at late times. This can be understood as follows: at early times, the stochastic method overpredicts the tunneling rate, as we saw in earlier sections. However, at late times, once the quantum particle escapes, it converts all its initial energy into kinetic energy and then moves out rapidly. While for the stochastic method, in most of the ensembles the particle gets trapped, while only occasionally it escapes. So the ensemble average is relatively suppressed at late times. These differences become even more extreme for larger $x_0$.

Therefore, in the simplest quantum mechanics example of single particle escape, exact Gaussian initial conditions are not sufficient to bring the stochastic approach into agreement with the true quantum evolution. Thus we attribute the discrepancy in the result to the procedure of classically evolving a quantum state in the stochastic method. It would be very informative to have exact simulations of quantum field theory to compare to the stochastic approach in that important context.

\section{Conclusions}\label{CN}

We quantitatively investigated the precision of the stochastic approach to tunneling in quantum field theory, comparing to the results of the standard instanton approximation. When the initial conditions on field fluctuations are drawn from the free quadratic theory Gaussian distributions and the evolution is then studied classically, we observed a significant quantitative disagreement  between the two methods, in contrast with the recent claim of excellent agreement in Ref.~\cite{Braden:2018tky}.\footnote{After the first version of our paper appeared, the authors of Ref.~\cite{Braden:2018tky} informed us that they indeed had an error; there was an accidental re-scaling of the parameter $\phi_0$ in their results. Hence, by their eq.(5), this is effectively equivalent to accidentally suppressing the fluctuation amplitudes.} It often does, however, have parametric agreement in the logarithm of the tunneling rate, as explained in other works \cite{Ellis:1990bv,Linde:1991sk,Hertzberg:2019wgx}. 

In particular, we find that in quantum field theory in $1+1$ dimensions the stochastic method overestimates the tunneling rate as compared to the instanton approximation. We showed that one needs to suppress fluctuations by an appropriate fudge factor $\fudge<1$ in order for the methods to be in good agreement. However, this comes at the expense of considering initial states which violate the uncertainty principle and therefore cannot arise from a Wigner distribution associated with a state in the Hilbert space. Also, we found that the optimal value of the fudge factor to mimic the instanton result depends on the shape and parameters of the potential under investigation. Furthermore, we found that for any choice of initial fluctuations that satisfy the uncertainty principle, and can therefore arise from the Wigner distribution of a physical state in the Hilbert space, the tunneling rate always far exceeds the instanton result. If the instanton is indeed accurately describing the quantum theory, then this means the stochastic approach is only parametrically correct in the logarithm of rates.

Note that this is all in contrast to the case in which the relevant modes begin at high occupancy; here the stochastic method of ensemble averaging from an initial positive definite Wigner distribution matches the quantum theory's predictions, as shown in Ref.~\cite{Hertzberg:2016tal}.

We made a first step towards comparing the predictions of the stochastic approach to the exact underlying quantum theory and testing the assumption of Gaussian initial conditions, by numerically solving the Schr\" odinger equation for a single particle quantum mechanics escape problem. We found that disagreement occurs both at early and late times. 

There are several directions for future work. If the goal is to bring the stochastic approach into alignment with the instanton approximation, then future work would be to develop a theory behind the optimal fudge factor $\fudge$. It may well be that the optimal value should be a function of $k$, i.e., $\fudge\to\fudge_k$. Furthermore, it would be interesting to understand if better quantitative agreement between these methods can be obtained by considering more than one spatial dimension, or special types of potentials, or in multi-field models. Other work would be to find exact tunneling rates in special quantum field theories, especially in regimes in which the instanton approximation is inaccurate, and then to directly test the predictions of the stochastic approach in such regimes.

\section*{Acknowledgments}

We thank Jose Blanco-Pillado, Larry Ford, Alan Guth, Mudit Jain, Ken Olum, Jiro Soda, Lorenzo Sorbo, Alex Vilenkin, Shao-Jiang Wang, and Masaki Yamada for helpful discussion. We also thank Francesc Ferrer and Jan Oll\' e for help with numerical integration in Python. M.~P.~H. is supported in part by National Science Foundation Grant No. PHY-2013953.

\end{document}